\documentclass[aip,jcp,reprint]{revtex4-2}

\usepackage[utf8]{inputenc}
\usepackage{amsmath}
\usepackage{amssymb}
\usepackage{graphicx}
\usepackage{hyperref}

\begin{document}

\title{Geometric Diagnostics of Scrambling-Related Sensitivity in a Bohmian Preparation Space}

\author{Stephen Wiggins}
\affiliation{Hetao Institute of Mathematics and Interdisciplinary Sciences, Shenzhen, China}
\affiliation{School of Mathematics, University of Bristol, UK}

\date{\today}

\begin{abstract}
The Out-of-Time-Order Correlator (OTOC) is a standard algebraic diagnostic of quantum information scrambling, but it offers limited direct geometric intuition. In this note, we propose a Bohmian, trajectory-based framework for constructing a geometric diagnostic of scrambling-related sensitivity using Lagrangian Descriptors (LDs). To avoid the uncertainty-principle obstruction to assigning independent initial position and momentum within a single wave function, we evaluate Bohmian dynamics over a two-dimensional preparation space of localized Gaussian wavepackets labeled by their initial center and momentum kick. For the inverted harmonic oscillator, this construction is analytically tractable: the wavepacket-center dynamics and their dependence on preparation parameters can be written explicitly. In particular, away from the equilibrium origin, the exponential growth of the associated preparation-space stability matrix yields an $\mathcal{O}(e^{\omega T})$ bound on the sensitivity of the wavepacket-center LDs, motivating a semiclassical comparison with sensitivity structures associated with OTOC growth. In this sense, the LD provides a geometric indicator of scrambling-related sensitivity. We conclude by discussing how this preparation-space picture suggests a program for future work regarding the distinct microcanonical regimes previously reported for the inverted harmonic oscillator.
\end{abstract}

\keywords{Quantum Scrambling, Out-of-Time-Order Correlator (OTOC), Lagrangian Descriptors, Bohmian Mechanics, Phase Space Geometry, Inverted Harmonic Oscillator}

\maketitle

\section{Introduction}
The study of quantum chaos and information scrambling has experienced a profound resurgence across multiple disciplines, from condensed matter physics to quantum gravity. The central algebraic diagnostic driving this resurgence is the Out-of-Time-Order Correlator (OTOC) \cite{larkin1969, kitaev2014, kitaev2018}. Originally introduced in the context of superconductivity, the OTOC has become the standard measure for the quantum ``butterfly effect,'' quantifying the rapid, exponential spread of localized quantum information across a system's degrees of freedom. In appropriate semiclassical or thermal settings, the growth rate of the OTOC defines a quantum Lyapunov exponent, famously bounded by fundamental thermodynamic constraints \cite{maldacena2016}.

However, because the OTOC is a fundamentally algebraic construct evaluated via non-commuting operators over abstract Hilbert spaces, it offers limited geometric intuition. In classical dynamical systems, chaos and information mixing are inherently visual, geometric phenomena. The classical butterfly effect is dictated by the stable and unstable invariant manifolds attached to hyperbolic saddle points. These manifolds form the impenetrable phase-space ``skeleton'' that governs transport and scattering---a conceptual framework utilized extensively in geometric Transition State Theory (TST) to understand reaction dynamics across energy barriers \cite{uzer2002, waalkens2004}. Recently, this chemical physics perspective has been unified with quantum information theory, revealing that barrier crossing and tunneling (instanton) dynamics at isolated phase-space saddles are fundamentally tied to the thermal quenching of the OTOC and the saturation of the quantum scrambling bound \cite{althorpe2023, althorpe2024, wolynes2024}.

To map these intricate phase-space structures in the processes generated by arbitrary time-dependent classical vector fields, Lagrangian Descriptors (LDs) were developed as a robust, trajectory-based scalar diagnostic \cite{madrid2009, mancho2013, lopesino2017}. By integrating the Euclidean arc-length of trajectories over a finite time window, LD ridges seamlessly isolate the invariant manifolds driving chaotic transport.

In this paper, we introduce a Bohmian preparation-space version of the Lagrangian Descriptor and analyze it for localized Gaussian states in the inverted harmonic oscillator. Our aim is not to define a new quantum observable, but rather to construct a geometric diagnostic associated with a family of prepared quantum states. Section 2 formulates the preparation-space framework and defines the forward and backward wavepacket-center LDs. Section 3 works out the Gaussian inverted harmonic oscillator dynamics explicitly, distinguishing the exact wavepacket-center evolution from the internal Bohmian motion within a wavepacket. Section 4 explains how the corresponding preparation-space stability matrix organizes wavepacket-center LD sensitivity and motivates, at a semiclassical level, a comparison with OTOC growth. Finally, Section 5 briefly discusses a conjectural extension of this picture to the microcanonical regimes identified in earlier exact OTOC calculations \cite{hashimoto2020}.

\section{Theoretical Framework}
In the Bohmian formulation of quantum mechanics, the wave function associated with a specific preparation $\mathbf{x}_0 = (q_0, p_0)$ is expressed in polar form as $\psi(q,t; \mathbf{x}_0) = R(q,t; \mathbf{x}_0) \exp(i S(q,t; \mathbf{x}_0)/\hbar)$, where $R$ is the real amplitude and $S$ is the real phase. Substituting this form into the time-dependent Schrödinger equation yields the quantum Hamilton-Jacobi equation:
\begin{equation}
\partial_t S + \frac{(\partial_q S)^2}{2m} + U(q) + Q(q,t; \mathbf{x}_0) = 0
\end{equation}
where $U(q)$ is the classical scalar potential and $Q(q,t; \mathbf{x}_0) = -\frac{\hbar^2}{2m} \frac{\partial_q^2 R}{R}$ is the quantum potential \cite{bohm1952}. 

Unlike classical mechanics, where position and momentum are independent phase-space coordinates, the Bohmian momentum is strictly constrained by the phase gradient, $p = \partial_q S(q,t; \mathbf{x}_0)$. The fundamental Bohmian object is therefore a configuration-space velocity field parameterized by the preparation:
\begin{equation}
\dot{q}_B = v_B(q,t; \mathbf{x}_0) = \frac{1}{m}\partial_q S(q,t; \mathbf{x}_0)
\end{equation}
By differentiating the quantum Hamilton-Jacobi equation with respect to $q$ and evaluating the convective derivative $\frac{d}{dt} = \partial_t + \dot{q}_B \partial_q$ along a Bohmian trajectory $q_B(t)$, one obtains the momentum evolution:
\begin{equation}
\begin{split}
\dot{p}_B(t) &= -\partial_q \left( U(q) + Q(q,t; \mathbf{x}_0) \right)\Big|_{q=q_B(t)}, \\
p_B(t) &= \partial_q S(q_B(t), t; \mathbf{x}_0)
\end{split}
\end{equation}
Thus, the pair $(q_B, p_B)$ does not constitute a globally defined vector field on an independent $(q,p)$-plane, but rather an induced first-order system along the evolving graph $p = \partial_q S(q,t; \mathbf{x}_0)$.

Because of this restriction, we cannot explore a full 2D phase-space grid of $(q_0, p_0)$ values using just one wave function. To overcome this, we must evaluate an \textit{ensemble} of different initial wave functions. For every distinct coordinate $(q_0, p_0)$ on our grid, we initialize a unique wavepacket that is physically centered at $q_0$ and given an initial momentum ``kick'' of $p_0$. 

It is vital to clearly distinguish this ensemble approach from classical phase-space analysis. In standard dynamical systems, a 2D grid represents a true physical phase space. To recover a genuinely two-dimensional geometric arena in a Bohmian setting, we introduce a \textit{Bohmian preparation space} (operationally referring here to the two-parameter manifold of Bohmian Gaussian preparations, distinct from general measurement contexts in quantum foundations \cite{mehrafarin2005}): the parameter manifold of a chosen family of localized initial wavepackets $\{\psi_{q_0, p_0}\}$, where each point $(q_0, p_0)$ labels a distinct preparation with wavepacket center $q_0$ and mean momentum $p_0$. This space should not be confused with the physical phase space of a single Bohmian trajectory. For a fixed wave function, the Bohmian momentum is constrained by the phase, $p = \partial_q S$, and $(q, p)$ are therefore not independent initial data for one trajectory. In contrast, $(q_0, p_0)$ here label different prepared states.

In this sense, the construction is analogous in spirit to coherent-state and Gaussian wavepacket phase-space representations \cite{heller1975, mizrahi1984, littlejohn1986, pattanayak1994, brif1999, coughtrie2015}, where families of minimum-uncertainty states are indexed by phase-space-like parameters. Such Gaussian parameterizations provide the natural comparison class because they form the standard semiclassical backbone for evaluating quantum thermodynamic and scrambling bounds. Accordingly, we do not claim the notion of a state-labeled preparation manifold itself to be fundamentally new; rather, the novelty lies in its specific Bohmian use as a two-parameter geometric domain for Lagrangian-descriptor analysis and its proposed connection to scrambling diagnostics.

There are three distinct objects in the discussion that should not be conflated. First, the pair $(q_0, p_0)$ labels the \textit{prepared initial state}. Second, for a Gaussian wavepacket, one may track the \textit{wavepacket-center trajectory} $(q_c(t), p_c(t))$, which depends analytically on $(q_0, p_0)$. Third, within any fixed prepared wavepacket, one may also consider the \textit{Bohmian particle trajectory} $q_B(t)$ determined by the full phase gradient of that specific wavepacket. 

While one could seek Lagrangian-descriptor-type diagnostics based on the quantum trajectory dynamics within each prepared wavepacket, our primary interest here is different. In recent work, Grover and Keshavamurthy \cite{grover2026} noted in the ETMD setting that trajectory-based quantum dynamics may provide a natural route toward extending Lagrangian descriptors into the quantum regime. Their focus, however, is on ETMD trajectories and transport/concertedness in a multi-degree-of-freedom chemical system, whereas the present construction is a Bohmian preparation-space diagnostic built from the motion of Gaussian wavepacket centers. We therefore introduce the \textit{wavepacket-center flow} (while general time-dependent potentials generate a two-parameter process, the quadratic model in Section 3 strictly reduces to an autonomous flow \cite{wiggins2025}) as the reduced dynamics on the preparation manifold and define the corresponding wavepacket-center LD from that flow:
\begin{equation}
\begin{split}
\Phi_c^t : \mathbf{x}_0 &\mapsto \mathbf{x}_c(t; \mathbf{x}_0), \\
\mathbf{x}_0 &= (q_0, p_0), \quad \mathbf{x}_c = (q_c, p_c),
\end{split}
\end{equation}
where $\mathbf{x}_c(t; \mathbf{x}_0)$ denotes the center trajectory associated with the preparation labeled by $\mathbf{x}_0$. The Lagrangian descriptor considered in this paper is then the descriptor associated with this reduced flow, rather than with the full Bohmian field inside each wavepacket. For brevity, we refer to this reduced descriptor as the \textit{wavepacket-center LD}.

Accordingly, we define the forward and backward wavepacket-center Lagrangian descriptors by:
\begin{align}
\mathcal{L}^{\text{wpc}}_{\text{fwd}}(\mathbf{x}_0, T) &= \int_{0}^{T} \|\dot{\mathbf{x}}_c(t; \mathbf{x}_0)\| dt \\
\mathcal{L}^{\text{wpc}}_{\text{bwd}}(\mathbf{x}_0, T) &= \int_{-T}^{0} \|\dot{\mathbf{x}}_c(t; \mathbf{x}_0)\| dt
\end{align}
where $\|\cdot\|$ denotes the standard Euclidean norm. To ensure dimensional consistency when summing position and momentum derivatives, the phase-space coordinates $\mathbf{x}_c$ are assumed to be suitably non-dimensionalized. To combine the forward and backward information into a single scalar diagnostic, we set:
\begin{equation}
\mathcal{M}^{\text{wpc}}(\mathbf{x}_0) = -\log_{10}\left( \mathcal{L}^{\text{wpc}}_{\text{fwd}}(\mathbf{x}_0, T) \times \mathcal{L}^{\text{wpc}}_{\text{bwd}}(\mathbf{x}_0, T) \right)
\end{equation}
Here, the base-10 logarithm compresses the massive exponential scale of the trajectory divergence into a viewable linear scale. Furthermore, because trajectories initialized exactly on the stable or unstable manifolds of a saddle do not diverge exponentially, the raw LD arc-length values are minimized along these structures. The negative sign inverts these ``valleys'' into sharp visual ``ridges,'' ensuring the classical skeleton stands out as a set of maxima in the diagnostic. This object should be understood as a preparation-space LD built from wavepacket-center trajectories.

\section{Scrambling at the Quantum Saddle: Explicit Trajectories}
To apply the ensemble framework developed in Section 2, we require a system where the wavepacket dynamics can be calculated precisely for any arbitrary initial condition. The inverted harmonic oscillator (IHO), defined by the classical potential energy $U(q) = -\frac{1}{2}m\omega^2q^2$, serves as the foundational toy model for a phase-space saddle. 

For each grid point $(q_0, p_0)$, we prepare a Gaussian wavepacket physically centered at $q_0$ with an initial momentum kick of $p_0$ \cite{barton1986}:
\begin{equation}
\psi(q,0; q_0, p_0) = \frac{1}{(2\pi \sigma_0^2)^{1/4}} \exp\left( -\frac{(q-q_0)^2}{4\sigma_0^2} + \frac{i p_0 q}{\hbar} \right)
\end{equation}
Because the classical potential $U(q)$ is strictly quadratic, an initial Gaussian wavepacket maintains a perfectly Gaussian spatial profile for all time. The variance $\sigma_t^2$ spreads deterministically \cite{barton1986}:
\begin{equation}
\sigma_t^2 = \sigma_0^2 \cosh^2(\omega t) + \frac{\hbar^2}{4 m^2 \omega^2 \sigma_0^2} \sinh^2(\omega t)
\end{equation}

Because the potential is strictly quadratic, the quantum dynamics of the wavepacket center decouple completely from its internal spreading. This allows us to state the following:

\textbf{Proposition 1:} \textit{For the Gaussian family in the quadratic IHO, the wavepacket center evolves exactly according to classical hyperbolic equations of motion, and the preparation-space Jacobian of this wavepacket-center flow is the classical hyperbolic matrix.}

\textit{Justification.} The classical hyperbolic evolution of the center is a direct consequence of Ehrenfest's theorem for strictly quadratic potentials \cite{heller1975, littlejohn1986}. The statement regarding the Jacobian then follows immediately by differentiating the resulting exact solutions (Eqs. \ref{eq:qc}--\ref{eq:pc}) with respect to the initial preparation parameters $(q_0, p_0)$. 

The central position $q_c(t)$ and central momentum $p_c(t)$ evolve analytically as:
\begin{align}
q_c(t) &= q_0 \cosh(\omega t) + \frac{p_0}{m\omega} \sinh(\omega t) \label{eq:qc} \\
p_c(t) &= m\omega q_0 \sinh(\omega t) + p_0 \cosh(\omega t) \label{eq:pc}
\end{align}
By separating the exact time-dependent Schrödinger equation solution for this spreading Gaussian into its polar form $\psi(q,t; q_0, p_0) = R \exp(i S/\hbar)$, the reconstructed real phase takes the exact form:
\begin{equation}
S(q,t; q_0, p_0) = p_c(t)(q - q_c(t)) + \frac{m \dot{\sigma}_t}{2\sigma_t} (q - q_c(t))^2 + \gamma(t) \label{eq:phase}
\end{equation}
where $\gamma(t)$ is a purely time-dependent global phase term. Because the Bohmian velocity field is determined strictly by the spatial gradient $\partial_q S$, this term vanishes upon differentiation and has no effect on the trajectory dynamics.

Differentiating Eq. (\ref{eq:phase}) with respect to $q$, the exact Bohmian configuration-space velocity field for this preparation is:
\begin{equation}
\dot{q}_B = v_B(q,t; q_0, p_0) = \frac{p_c(t)}{m} + \frac{\dot{\sigma}_t}{\sigma_t}(q - q_c(t)) \label{eq:vb}
\end{equation}
This explicit formula clearly distinguishes two notions of instability in the present problem. The first term, $p_c(t)/m$, corresponds to the exact hyperbolic evolution of the wavepacket center $(q_c(t), p_c(t))$, which for the quadratic inverted oscillator is purely classical and is given by Eqs. (\ref{eq:qc})--(\ref{eq:pc}). The second term represents the local Bohmian deformation within the wavepacket, which is proportional to the spreading rate $\dot{\sigma}_t / \sigma_t$. This internal deformation may modify local stretching of Bohmian trajectories inside a wavepacket, even though the center evolution itself remains exactly classical.

For narrow initial wavepackets, the uncertainty principle forces a large initial momentum spread, which in Bohmian language is encoded in a large initial quantum potential. This affects the local internal structure of the Bohmian velocity field through the time dependence of $\sigma_t$. However, for the strictly quadratic inverted oscillator considered here, the wavepacket-center dynamics remain exactly those of the classical saddle. Accordingly, any preparation-space stability matrix built from the center evolution inherits the classical hyperbolic rate $\omega$, while quantum effects enter through the internal wavepacket structure and through the interpretation of localized prepared states.

When we evaluate the wavepacket-center diagnostic $\mathcal{M}^{\text{wpc}}(\mathbf{x}_0)$ over the preparation grid, the resulting ridges identify the stable and unstable organization of the preparation-space dynamics. In the quadratic inverted oscillator, because the wavepacket-center evolution is exactly classical, this diagnostic recovers the classical saddle skeleton associated with the wavepacket-center flow. 

To be completely explicit: because the IHO potential is strictly quadratic, the wavepacket-center LD computed here does not demonstrate a distinctly new quantum scrambling mechanism. Instead, it provides an exact reinterpretation within this reduced wavepacket-center construction of the classical saddle geometry evaluated over a quantum-preparation ensemble. The quantum nature of the system is encoded in the dimensionality and uncertainty bounds of this preparation space, rather than in a non-classical deformation of the central saddle skeleton.

\section{Semiclassical Heuristics for the OTOC}
The OTOC is the standard algebraic diagnostic of scrambling, whereas the Lagrangian Descriptor is a geometric diagnostic of trajectory sensitivity. Our goal in this section is not to identify these objects exactly, but to explain why they are naturally organized by a common stability structure in semiclassical settings. 

For the quadratic inverted harmonic oscillator, the wavepacket-center dynamics uniquely define an autonomous linear flow on preparation space. Writing $\mathbf{x}_c(t; \mathbf{x}_0) = \Phi_c^t(\mathbf{x}_0)$, the evolution is generated by:
\begin{equation}
\dot{\mathbf{x}}_c = \mathbf{f}_c(\mathbf{x}_c), \quad \mathbf{f}_c(\mathbf{x}) = A\mathbf{x}
\end{equation}
with the constant matrix:
\begin{equation}
A = \begin{pmatrix} 0 & 1/m \\ m\omega^2 & 0 \end{pmatrix}
\end{equation}
Hence, the center trajectory is given by $\mathbf{x}_c(t; \mathbf{x}_0) = e^{At}\mathbf{x}_0$, and the Jacobian of this wavepacket-center flow with respect to the preparation parameters is exactly the classical hyperbolic matrix:
\begin{equation}
\mathbf{J}_c(t) = D_{\mathbf{x}_0}\Phi_c^t(\mathbf{x}_0) = e^{At} = \begin{pmatrix} \cosh(\omega t) & \frac{1}{m\omega} \sinh(\omega t) \\ m\omega \sinh(\omega t) & \cosh(\omega t) \end{pmatrix}
\end{equation}
In the present quadratic model, the explicit Jacobian used below is the Jacobian of the wavepacket-center flow, so its exponential rate is classical; quantum effects enter elsewhere in the Bohmian structure and in the interpretation of the preparation ensemble. 

\textbf{1. Relation to scrambling diagnostics.}
In semiclassical settings, the connection between OTOCs and classical trajectory sensitivity is already well established. For canonical choices of observables, the squared commutator reduces at leading order to a squared Poisson bracket evaluated along the classical trajectory. Because Proposition 1 establishes that the classical trajectory in this model is exactly the wavepacket-center trajectory $\mathbf{x}_c(t)$, we can write the semiclassical OTOC explicitly in terms of our preparation parameters:
\begin{equation}
C(t) \simeq \hbar^2 \{q_c(t), p_c(0)\}^2 = \hbar^2 \left( \frac{\partial q_c(t)}{\partial q_0} \right)^2
\end{equation}
which corresponds exactly to a squared element of the preparation-space stability matrix $\mathbf{J}_c(t)$ \cite{jalabert2018}. Subsequent work has extensively demonstrated that semiclassical OTOC growth is governed by classical Lyapunov exponents and local stability matrices, particularly for states localized near unstable fixed points \cite{hummel2019, xu2020, steinhuber2023, meier2023}. Therefore, our aim here is not to establish the OTOC--stability matrix connection itself. Rather, we propose that this established sensitivity mechanism can be geometrically mapped using Lagrangian Descriptors evaluated over a Bohmian preparation space---a formulation that, rather than proposing a new scrambling mechanism, provides a Bohmian preparation-space geometric representation of sensitivity structures that arise semiclassically in scrambling diagnostics.

\textbf{2. Relation to Lagrangian Descriptors.}
The wavepacket-center LD depends on the reduced flow field $\mathbf{f}_c$ through the trajectory integral defining $\mathcal{L}^{\text{wpc}}_{\text{fwd}}$. Differentiating with respect to the preparation parameters introduces the Jacobian $\mathbf{J}_c(t)$ through the chain rule. While this type of gradient bound is the standard mathematical mechanism used to explain ridge formation in classical phase-space LDs \cite{mancho2013, lopesino2017}, its application here evaluates sensitivity with respect to the quantum preparation manifold rather than a classical state space.

\textbf{Proposition 2 (Asymptotic gradient bound).} \textit{Let
\begin{equation}
\mathcal{L}^{\text{wpc}}_{\text{fwd}}(\mathbf{x}_0, T) = \int_0^T \|\mathbf{f}_c(\mathbf{x}_c(t; \mathbf{x}_0))\| dt
\end{equation}
be the forward wavepacket-center Lagrangian descriptor. For $\mathbf{x}_0 \neq 0$ such that $\mathbf{f}_c(\mathbf{x}_c(t; \mathbf{x}_0)) \neq 0$ on $0 \le t \le T$, its sensitivity with respect to the preparation parameters satisfies
\begin{equation}
\|\nabla_{\mathbf{x}_0} \mathcal{L}^{\text{wpc}}_{\text{fwd}}(\mathbf{x}_0, T)\| \le \int_0^T \|D\mathbf{f}_c(\mathbf{x}_c(t; \mathbf{x}_0))\| \|\mathbf{J}_c(t)\| dt
\end{equation}
In particular, for the quadratic inverted harmonic oscillator away from the equilibrium origin, $\|\nabla_{\mathbf{x}_0} \mathcal{L}^{\text{wpc}}_{\text{fwd}}(\mathbf{x}_0, T)\| = \mathcal{O}(e^{\omega T})$.}

\textit{Proof.} Since $\mathbf{x}_c(t; \mathbf{x}_0) = \Phi_c^t(\mathbf{x}_0)$, we have $\mathcal{L}^{\text{wpc}}_{\text{fwd}}(\mathbf{x}_0, T) = \int_0^T \|\mathbf{f}_c(\mathbf{x}_c(t; \mathbf{x}_0))\| dt$. Differentiating with respect to $\mathbf{x}_0$ (which is valid away from the origin where the Euclidean norm is smooth) and applying the chain rule gives the column gradient vector:
\begin{equation}
\nabla_{\mathbf{x}_0} \mathcal{L}^{\text{wpc}}_{\text{fwd}} = \int_0^T \mathbf{J}_c(t)^T D\mathbf{f}_c(\mathbf{x}_c(t; \mathbf{x}_0))^T \frac{\mathbf{f}_c(\mathbf{x}_c(t; \mathbf{x}_0))}{\|\mathbf{f}_c(\mathbf{x}_c(t; \mathbf{x}_0))\|} \, dt
\end{equation}
Taking Euclidean norms yields:
\begin{equation}
\|\nabla_{\mathbf{x}_0} \mathcal{L}^{\text{wpc}}_{\text{fwd}}\| \le \int_0^T \|D\mathbf{f}_c(\mathbf{x}_c(t; \mathbf{x}_0))\| \|\mathbf{J}_c(t)\| dt
\end{equation}
For the quadratic inverted harmonic oscillator, $\mathbf{f}_c(\mathbf{x}) = A\mathbf{x}$, so $D\mathbf{f}_c = A$ is a constant matrix. Therefore,
\begin{equation}
\|\nabla_{\mathbf{x}_0} \mathcal{L}^{\text{wpc}}_{\text{fwd}}\| \le \|A\| \int_0^T \|\mathbf{J}_c(t)\| dt
\end{equation}
Since $\mathbf{J}_c(t) = e^{At}$ has hyperbolic growth rate $\omega$, we obtain $\int_0^T \|\mathbf{J}_c(t)\| dt = \mathcal{O}(e^{\omega T})$, and hence:
\begin{equation}
\|\nabla_{\mathbf{x}_0} \mathcal{L}^{\text{wpc}}_{\text{fwd}}(\mathbf{x}_0, T)\| = \mathcal{O}(e^{\omega T})
\end{equation}
This shows that the preparation-space sensitivity of the forward wavepacket-center LD is bounded above by the hyperbolic growth scale that governs the wavepacket-center flow. By time-reversal symmetry, an identical $\mathcal{O}(e^{\omega T})$ bound governs the backward descriptor $\mathcal{L}^{\text{wpc}}_{\text{bwd}}(\mathbf{x}_0, T)$, with its sensitivity localized along the complementary invariant manifold. Consequently, large preparation-space stretching amplifies the sensitivity of both LDs to $(q_0, p_0)$ at a rate bounded by $\mathcal{O}(e^{\omega T})$, providing a mechanism consistent with the sharpening of LD ridges near organizing structures of the flow.

Thus, although the LD is not an OTOC and the preparation-space Jacobian is not identical to an exact quantum commutator, both are organized by sensitivity of the underlying evolution to initial data. In the analytically tractable setting considered here, the LD ridges may therefore be interpreted as a geometric indicator of scrambling-related sensitivity in preparation space. 

\begin{figure}[htbp]
    \centering
    \includegraphics[width=\columnwidth]{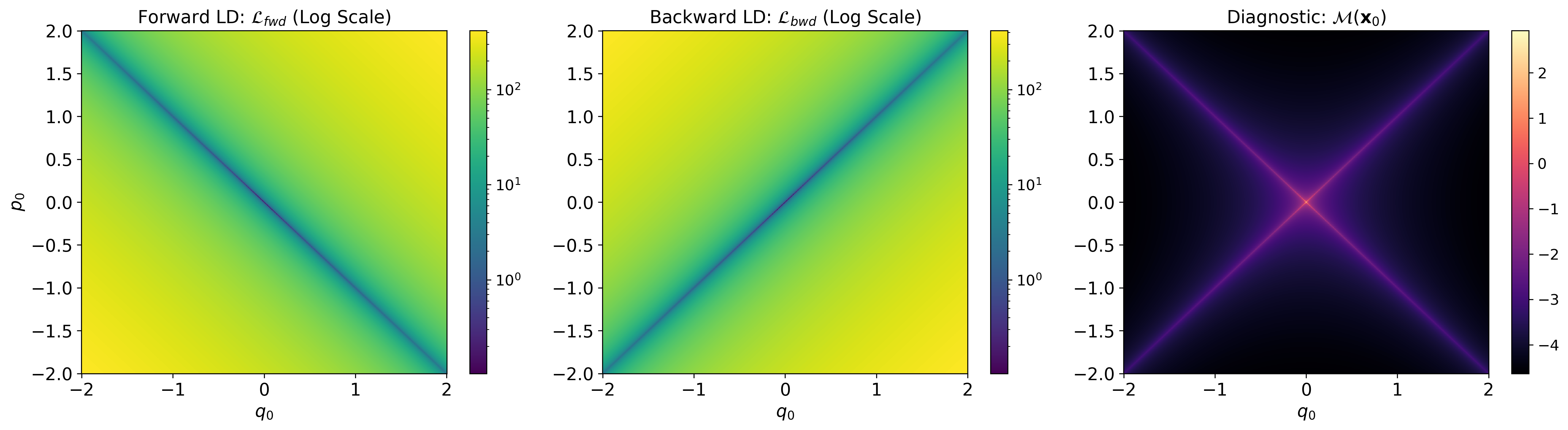}
    \caption{Wavepacket-center Lagrangian Descriptors for the preparation-space dynamics of the inverted harmonic oscillator ($m=1, \omega=1$). The diagnostic is evaluated over a $1000 \times 1000$ grid of initial preparation parameters $(q_0, p_0)$ with an integration horizon of $T=5$. (Left and Center) The forward (Eq. 5) and backward (Eq. 6) wavepacket-center LDs, plotted on a logarithmic scale. Because LD arc-lengths are minimized on the invariant manifolds of a saddle, the stable and unstable manifolds appear as dark trenches. (Right) The product diagnostic $\mathcal{M}^{\text{wpc}}(\mathbf{x}_0)$ (Eq. 7). The negative logarithm compresses the exponential sensitivity and inverts the LD minima, revealing the underlying classical wavepacket-center skeleton as bright ridges in preparation space.}
    \label{fig:LD_manifolds}
\end{figure}

\textbf{Numerical Illustration.} 
The results displayed in Fig.~\ref{fig:LD_manifolds} visually illustrate the relationship between preparation-space sensitivity and the stability matrix $\mathbf{J}_c(t)$. By increasing the integration horizon to $T=5$, the exponential stretching factor grows to $e^{\omega T} \approx 148$. On the logarithmic scale (Fig.~\ref{fig:LD_manifolds}, left/center), the LD values vary across three orders of magnitude, with the steepest gradients localized precisely along the invariant manifolds. This is consistent with the interpretation that the LD ridges act as a geometric indicator for the directions of maximal sensitivity in preparation space. 

\section{Outlook: Conjectural Extension to Microcanonical Regimes}
Up to this point, the Lagrangian Descriptor has been defined over a preparation space of localized wavepackets. This differs conceptually from the microcanonical setting in which the OTOC is evaluated in specific energy eigenstates. For the inverted harmonic oscillator, Hashimoto et al. \cite{hashimoto2020} computed the exact expectation value $c_n(t) = -\langle n | [\hat{q}(t), \hat{p}(0)]^2 | n \rangle$ and identified distinct temporal regimes depending on the state. 

While we do not derive the Bohmian hydrodynamics of those scattering or eigenstate-like solutions here, the geometric preparation-space picture developed in this note suggests a natural program for future work. Specifically, one may hypothesize that state-dependent quantum potentials could induce qualitatively different effective geometries, and that these geometries might help organize the different scrambling behaviors seen in exact microcanonical OTOC calculations:

\begin{itemize}
    \item \textbf{Low-energy / deep-tunneling regime:} One possible scenario is that states deep below the barrier effectively suppress hyperbolic instability. The corresponding diagnostic would then show weak ridge formation, qualitatively consistent with non-exponential or oscillatory scrambling signals.
    \item \textbf{Near-barrier regime:} One may expect that states concentrated near the barrier top retain the strongest hyperbolic character, making them the primary candidates for displaying pronounced ridge formation and rapid scrambling-like growth.
    \item \textbf{High-energy / over-the-barrier regime:} It is plausible that for highly excited states, the relevant motion spends too little time near the unstable region for substantial sensitivity to accumulate. The associated geometric signature would then flatten, in qualitative agreement with weak net scrambling growth.
\end{itemize}

A genuine test of this conjectural framework would require constructing the exact Bohmian processes associated with the relevant inverted-oscillator scattering states and comparing their stability structure directly with the exact microcanonical OTOCs.

\section{Conclusions}
In this note, we introduced a Bohmian preparation-space framework for constructing a Lagrangian-Descriptor-based geometric diagnostic of scrambling-related sensitivity. For localized Gaussian states in the inverted harmonic oscillator, the dependence of the dynamics on the preparation parameters can be written explicitly, making the associated stability structure analytically transparent. Specifically, we demonstrated that the exponential growth of the preparation-space stability matrix yields an analytical $\mathcal{O}(e^{\omega T})$ bound on the sensitivity of the forward and backward wavepacket-center LDs. This provides a natural geometric lens on sensitivity to initial quantum preparation, and it suggests a semiclassical relation between LD ridge formation and the mechanisms usually associated with OTOC growth. At the same time, the present analysis remains limited to a tractable model and to a preparation-space construction rather than an exact identification with the OTOC itself. A fuller quantitative connection, especially in eigenstate-based or genuinely nonquadratic settings, remains an important problem for future work.

\end{document}